\documentclass{aa}  

\usepackage{graphicx}
\usepackage{txfonts}
\usepackage[dvipsnames]{xcolor}
\usepackage[utf8]{inputenc}
\usepackage{ulem}

\newcommand{\teff}{$T_{\mathrm{eff}}$}
\newcommand{\kms}{km~s$^{-1}$}
\newcommand{\msun}{M$_\odot$}
\newcommand{\logg}{$\log g$}
\newcommand\omicron{o}
\newcommand{\omiori}{$\omicron^1$~Ori}
\newcommand{\BD}{BD+79$^\circ$156}

\newcommand{\AJ}[1]{\textcolor{black}{#1}} 
\newcommand{\LS}[1]{\textcolor{black}{#1}}  
\newcommand{\SVE}[1]{\textcolor{black}{#1}}  

\begin{document}

\title{Discovery of technetium- and niobium-rich S stars: the case for bitrinsic stars \thanks{Based on observations made with the Mercator Telescope, operated on the Island of La Palma by the Flemish Community, at the Spanish {\it Observatorio del Roque de los Muchachos} of the {\it Instituto de Astrofísica de Canarias}.}}

   \author{S. Shetye
          \inst{1,2}
          \and
          S. Van Eck\inst{1} \and
          S. Goriely\inst{1} \and
          L. Siess\inst{1}   \and
          A. Jorissen \inst{1} \and
          A. Escorza \inst{2,1}\and
          H. Van Winckel\inst{2}
          }

   \institute{ Institute of Astronomy and Astrophysics (IAA), Université libre de Bruxelles (ULB)
,
              CP  226,  Boulevard  du  Triomphe,  B-1050  Bruxelles, Belgium\\
              \email{Shreeya.Shetye@ulb.ac.be}
         \and
             Institute of Astronomy, KU Leuven, Celestijnenlaan 200D, B-3001 Leuven, Belgium
             }

   \date{Received; accepted }

 
  \abstract
   {\LS{S stars are late-type giants with overabundances of s-process elements. They come in two flavours depending on the presence or not of technetium (Tc), an element without stable isotopes. Intrinsic S stars are Tc-rich and genuine asymptotic giant branch (AGB) stars while extrinsic S stars owe their s-process {\SVE{ over}}abundances to the pollution from a former AGB companion, now a white dwarf (WD). In addition to Tc, another distinctive feature  between intrinsic and extrinsic S stars is the overabundance of niobium (Nb) in the latter class. }
   \SVE{Indeed, since the mass transfer occurred long ago, $^{93}$Zr had time to decay into the only stable isotope of Nb,  $^{93}$Nb, causing its overabundance.}
   }
   {\LS{We discuss the case of the S stars \BD{} and \omiori{}{} whose} \AJ{specificity is to share the distinctive features of both intrinsic and extrinsic S stars, namely the presence of Tc along with a Nb overabundance. }}
   {We used high-resolution HERMES optical spectra, MARCS model atmospheres of S stars, Gaia DR2 parallaxes and STAREVOL evolutionary tracks to determine the stellar parameters and chemical abundances of the two S stars, and to locate them in the Hertzsprung-Russell (HR) diagram. } 
   {\BD{} is the first \AJ{clear} case of a  bitrinsic star, i.e. a doubly s-process-enriched object, \AJ{first through mass transfer in a binary system,
   and then through internal nucleosynthesis (responsible for the 
   \SVE{Tc-enrichment in} \BD{} which must therefore have reached the AGB phase of its evolution). This hybrid nature of the s-process pattern in \BD{} is supported by its binary nature and its location in the HR diagram just beyond the onset of the third dredge-up on the AGB.} 
   \AJ{The Tc-rich, binary S-star \omiori{} with a WD companion was another long-standing candidate for a similar hybrid s-process enrichment. 
   \SVE{However} the marginal overabundance of Nb derived in \omiori{} 
   \SVE{does not allow to trace unambiguously the evidence of a large pollution coming from the AGB progenitor of its current WD companion.}
   }
   \AJ{As a side product, the current study offers a new way of detecting binary AGB stars with WD companions by identifying  their Tc-rich nature along with a Nb overabundance.}
   }
   {}


   \maketitle
%

\section{Introduction}

S stars are late-type giants with molecular bands of ZrO and TiO as the most characteristic spectroscopic feature \citep{Merrill}. 
They are transition objects between M-type stars and carbon stars on the \AJ{asymptotic giant branch (AGB)} as their carbon/oxygen (C/O) ratio is in the range 0.5 $\leq$ C/O <~1 \citep{IbenRenzini,scaloross1976,sophieMarcs}. 
The strong ZrO bands actually reflect overabundances of slow neutron-capture (s-process) elements in their spectra \citep{keenan1954,smith1990}.

\LS{S stars come in two flavours. Intrinsic S stars  are defined as Tc-enriched S stars.} \AJ{Since Tc is an  element with no stable isotopes and its $^{99}$Tc isotope produced by the s-process has a half-life $T_{1/2}$ of $2.1\times10^5$ yr, its detection indicates that this element is currently synthesized by the star located on the AGB} \LS{and which is undergoing third-dredge-up (TDU) episodes. On the other hand, in extrinsic S stars, Tc was brought} \AJ{long ago by a former AGB companion through mass transfer. Because the time elapsed since this mass-transfer episode is most likely much longer than the $^{99}$Tc half-life,}
\LS{Tc had time to decay in $^{99}$Ru so that Tc lines are thus absent from the spectrum.
Extrinsic S stars that are not yet on the thermally-pulsing (TP-) AGB can thus not produce s-process elements themselves \citep{hipp, shreeya1}. 
They have all been shown to be binaries \citep{jorissen1998,jorissen2019}.}

\LS{The isotope $^{93}$Zr has a half-life $T_{1/2}= 1.53\times10^6$~yrs, and is another radio-isotope the abundance of which corroborates the above scenario. In extrinsic S stars, this isotope had time to  decay into  $^{93}$Nb, the only stable Nb isotope.
On the contrary, in intrinsic S stars, $^{93}$Zr is still alive, so that the Nb abundance is still low  \citep[{\it i.e.}, with a pre-s-process, solar-scaled abundance;][]{pieter2015}. }

In the current work, \LS{through abundance analyses of \BD{} and \omiori{}}, we present a new type of S star, 
showing signatures of both intrinsic and extrinsic S stars. We label them 'bitrinsic' S stars. 
\SVE{A bitrinsic star can be defined as a star having passed through a double pollution in s-process elements, first when its companion was on the TP-AGB and polluted it through mass transfer, and second when itself evolves on the TP-AGB and enriches its envelope by dredge-ups. These two events can only be traced by measuring Nb (enhanced in mass-transfer systems) and Tc (enhanced in TP-AGB objects) abundances.
}
This new class of S stars has been unraveled thanks to the availability of high-resolution spectra and of Gaia data-release 2 (GDR2) parallaxes \citep{gdr2}. \LS{In Sect.~\ref{Sect:observations}, we describe the observations, the detection of Tc and discuss the binary nature of the studied systems. In Sects.~\ref{Sect:atmosphericparameters}, \ref{Sect:abundances} and \ref{Sect:HRD}, we present the atmospheric parameters, surface abundances and location in the Hertzsprung-Russell (HR) diagram, which confirms the genuine AGB evolutionary status of these bitrinsic candidates. Abundances are then compared to  STAREVOL nucleosynthesis predictions 
in Sect.~\ref{SECT:STAREVOLpredictions}. Section~\ref{Sect:Conclusions} summarises our analysis.}


\section{Searching for bitrinsic S stars}\label{Sect:observations}

Among the S stars from the General Catalogue of Galactic S Stars \citep[CGSS]{cgss}, we selected a sample of S stars fulfilling the following quality conditions:
\begin{enumerate}
    \item Stars with $\sigma_{\varpi} / \varpi$~$\leq 0.3$, where $\varpi$ is the Gaia parallax and $\sigma_{\varpi}$ its error;
    \item Stars with declinations north of -30$^\circ$ and $V$ magnitudes brighter than 11~mag, to be observed with the HERMES high-resolution spectrograph\footnote{HERMES is a high-resolution ($R=86\,000$) spectrograph mounted on the Mercator telescope at the Roque de los Muchachos Observatory, La Palma (Canary islands).} \citep{raskin};
    \item Stars with HERMES spectra having a signal-to-noise ratio (SNR) of 50 in the $V$ band, since a high SNR is required to ensure the Tc-line detectability \citep{sophieMarcs, shreeya1}. 
\end{enumerate}

\LS{Out of this sample, we found two candidates, namely \BD{} and \omiori{}, that show the presence of Tc lines and Nb overabundances. The following sections present the detailed analysis of these stars.}
\SVE{The analysis of the remaining of this sample is presented elsewhere \citep{shreeya1, shreeya2} and in a forthcoming paper (Shetye et al., in prep.).}

\subsection{Technetium detection}\label{Sect:Tc}
\LS{The detection of Tc is made using the three \ion{Tc}{I} resonance lines located at 4238.19~\AA, 4262.27~\AA, and 4297.06~\AA.} Based on medium-resolution spectra, \cite{jorissen1993} classified \BD{} as a Tc-deficient object. 
 A subsequent study of infrared colors  \citep[which constitute an indirect classification criterion, since intrinsic stars generally exhibit infrared excesses indicative of mass loss;][]{jorissen1993, sophie2000} by \cite{yang2006} and \cite{otto2011} classified \BD{} as well as an extrinsic S star. Regarding \omiori{}, \cite{wallerstein1988} and \cite{smith1988} derived its Tc abundance from medium-resolution spectra and thus classified it as 
 intrinsic \SVE{(Tc-rich)}.
 This claim was confirmed by \cite{wang} from near-infrared observations. 
 
\begin{figure}[!htbp]     
\begin{centering}
\includegraphics[scale=0.5,trim={2cm 0cm 2cm 1cm},clip]{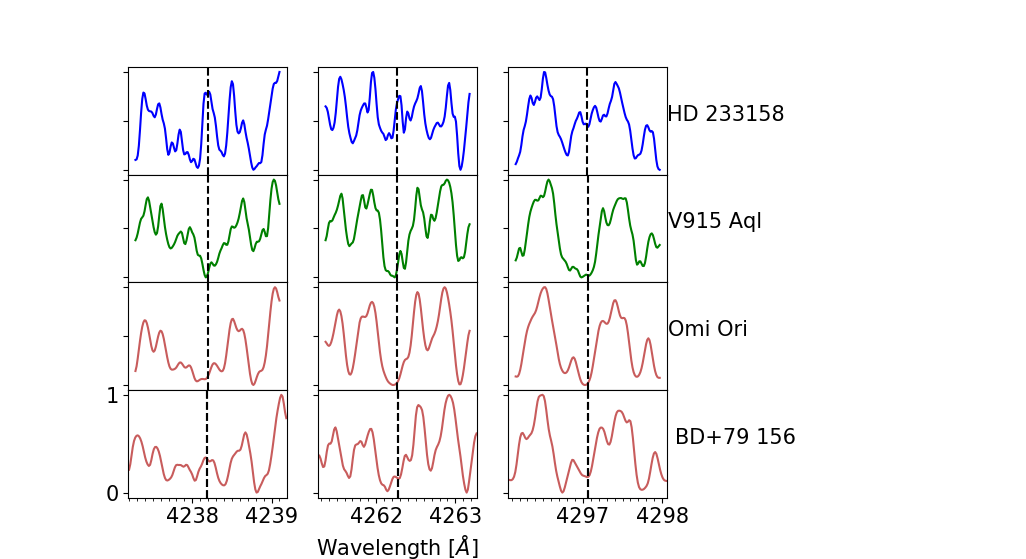}
    \caption{\label{Tc1}
    The spectral region around the three
    (4238.19, 4262.27 and 4297.06~\AA) ultra-violet \ion{Tc}{I} lines for Tc-poor (HD 233158, blue spectrum in the top panels) and Tc-rich (V915~Aql, green spectrum in the second from top panels) S stars from \cite{shreeya1} and for the bitrinsic candidates \omiori{} and \BD{} (bottom panels). The spectra have been arbitrarily normalized and binned by a factor of 1.5 to increase the SNR.}
\end{centering}
\end{figure}

\AJ{However, as shown in Fig.~\ref{Tc1} based on high-resolution HERMES spectra, absorption features are clearly present \SVE{at the wavelengths of the three UV Tc lines} for the two stars \omiori\ and  \BD{}. The 
\SVE{spectra }of 
\SVE{\BD{}, \omiori{} and of} a reference Tc-rich S star (V915 Aql)  
\SVE{are}
in sharp contrast with that of the
\SVE{reference} Tc-poor S star HD~233158, shown for comparison. From our analysis of high-resolution spectra, we thus re-classify \BD{} as a Tc-rich S star (Fig.~\ref{Tc1}). }

\subsection{Binarity}
\label{Sect:binarity}

\AJ{The binary nature of \omiori{} was first assessed by the discovery of a white dwarf (WD) companion in {\it International Ultraviolet Explorer} (IUE) data, with spectral type DA3 and \teff~$ = 22\,000$~K  \citep{OmiOriAke}. Prior to that discovery,
\cite{peery1986} had reported the detection of emission lines from ionised elements in IUE spectra of \omiori, but could not decide whether these lines originate from the giant star (due to flares for instance) or were attributable to the presence of a hot subluminous companion.    \omiori{} has since been an enigma 
because Tc-rich S stars  ought not to have a WD companion, in contrast to Tc-poor stars. But it has remained unclear until now whether this WD had polluted its companion with heavy elements while on the TP-AGB, or in other words,  whether this star was extrinsic before turning intrinsic. This question will be addressed in Sect.~\ref{SECT:STAREVOLpredictions}.  }

\AJ{The derivation of the orbital elements of the  \omiori{} system has turned out to be very difficult \citep{jorissen2019}, because the radial-velocity data show a very limited excursion of 2 to 3~\kms\ with some superimposed radial-velocity jitter. A tentative period of 575~d and a mass function of $(2.7 \pm 0.2) \times 10^{-5}$~M$_\odot$ was obtained by \citet{jorissen2019}, suggestive of an orbit seen almost pole-on ($i = 4.7^\circ\pm1^\circ$).}

\AJ{\BD{} is a confirmed binary with an orbital period of 10931 days, an eccentricity of 0.461, and a mass function of 0.025~\msun, compatible with a WD companion \citep{mathieu2017,jorissen2019}. An additional confirmation 
of the presence of a WD companion around \BD{} (and \omiori{}) comes from the UV excess seen in  their spectral energy distribution (SED), well visible when comparing photometric observations with model SED (Fig.~\ref{SEDs}). 
\SVE{
The binarity criteria proposed by \cite{ortizUV} based on the GALEX near-UV observed to predicted flux ratios
confirm that both \omiori{} and \BD{} are in a binary system.}}  

\section{Derivation of the atmospheric parameters}\label{Sect:atmosphericparameters}

The stellar parameters impacting the atmospheric thermal structure of S stars are \teff, \logg, [Fe/H], C/O and [s/Fe]. 
To disentangle this multi-dimensional parameter space, 
we follow the same methodology as described in \cite{shreeya1}. Briefly, this method performs a $\chi^2$ minimisation search over several spectral windows, comparing the observed spectrum with synthetic MARCS model atmospheres of S stars \citep{sophieMarcs}.
The best-fitting model is the one with the lowest $\chi^{2}$ value and the corresponding parameters 
constitute our first estimates.

The luminosity derived from the Gaia parallax
and the first estimate of \teff~from the $\chi^{2}$ fitting tool allow us to locate the star in the HR diagram. From this, we derive the stellar mass and hence the surface gravity by comparing this position in the HR diagram with STAREVOL evolutionary tracks. We then perform
an iteration on \logg\ and on the stellar parameters, as described in \cite{shreeya1}, to obtain a $\log g$ value consistent with the mass inferred from the location in the HR diagram.
\LS{The mass so derived is $\approx$ 2~\msun\ for \BD{} and 1.5~\msun\ for \omiori{}.}
The final list of stellar parameters is presented in Table~\ref{basicdataandparameters}\footnote{We note that the atmospheric parameters of \omiori{} quoted in \cite{jorissen2019} are different than the ones used in this study. This is because \logg~iterations were not performed in \cite{jorissen2019}.}. 
\begin{table}
    \caption{Main properties of the bitrinsic S stars. $V$ and $K$ magnitudes are extracted from the SIMBAD Astronomical Database \citep{simbad}. The parallax $\varpi$ and its uncertainty  $\sigma_{\varpi}$ are from  GDR2  \citep{gdr2}. The reddenning $E_{B-V}$ has been obtained from the \cite{gontcharov} extinction maps. $BC_K$ is the bolometric correction in the $K$ band as computed from the MARCS model atmospheres. Atmospheric parameters for bitrinsic stars are also given  (see Sect.~\ref{Sect:atmosphericparameters}). Values between parentheses correspond to the parameter range provided by the $\chi^2$ fitting. The metallicities [Fe/H] have been obtained from the  Fe abundance analyses. The  standard deviation $\sigma_{\rm [Fe/H]}$ represents the line-to-line scatter on [Fe/H]. The masses have been derived from the locations of the stars in the HR diagram compared to the STAREVOL tracks.}
    \label{basicdataandparameters}
    \centering
    \begin{tabular}{c|cc}
    \hline
          & \omiori{} & \BD{} \\
          \hline
          &\multicolumn{2}{c}{Basic data}\\
         \hline\\
         Sp. type & M3SIII & S4/2  \\
         $V$ (mag) & 4.72 & 9.55 \\
         $K$ (mag) & 0.48 & 5.67  \\
         $\varpi\pm\sigma_{\varpi}$ (mas) & 6.16 $\pm$ 0.41 & 0.57 $\pm$ 0.03\\
         $E_{B-V}$ & 0.07 & 0.26 \\
         $BC_{K}$ (mag) & 2.91 & 2.85 \\
         \hline
         &\multicolumn{2}{c}{Atmospheric parameters}\\
         \hline\\
         \teff~(K) & 3500 & 3600 \\
         & (3500 - 3600) & (3600 - 3700)  \\
         $L (L_\odot)$ & 2700 & 2800  \\
          & (2300 - 3100) & (2700 - 2900) \\
          \logg & 1 & 1 \\
          & (0 - 2) & (1 - 3) \\
          $M$ (\msun) & 1.5 & 2.0 \\
          & (1.3 - 2) & (1.5 - 2.5)  \\
          C/O & 0.75 & 0.50 \\
          & (0.50 - 0.90) & (0.50 - 0.75) \\
          $\rm[s/Fe]$ & 0 & 1  \\
          & (0 - 1) & (1 - 1) \\
          $\rm[Fe/H]$ & -0.28 & -0.16 \\
          $\sigma_{\rm [Fe/H]}$ & 0.19 & 0.12 \\
          \hline
    \end{tabular}
\end{table}

\section{Abundance determination}\label{Sect:abundances}
We used the radiative transfer code TURBOSPECTRUM \citep{rodrigo, turbospectrum} to generate synthetic spectra from the MARCS model atmospheres of the S stars with the atmospheric parameters determined as described in Sect.~\ref{Sect:atmosphericparameters}. 
We used the C/O estimates from the $\chi^2$ spectral fitting tool and checked
whether these values reproduce the observed spectra 
satisfactorily in the CH bands around 4200 \AA\ -- 4300 \AA.
Because atomic oxygen lines are poorly reproduced in S-type stars, we adopted the solar oxygen abundance from \cite{asplund2009}, scaled to the star metallicity. The nitrogen abundance was derived using the CN bands at 7900 \AA\ -- 8100 \AA.  

The metallicity was determined
using at least ten isolated Fe lines in the region 
from
7350 \AA\ to 8730 \AA. The derived [Fe/H] and its
standard deviation 
are provided
in Table~\ref{basicdataandparameters}. 
 The s-process abundances are listed in Table~\ref{abundancestrinsic}. 
 The Zr abundance was derived using two \ion{Zr}{I} lines at 7819.37 \AA\ and 7849.37~\AA\ with transition probabilities from laboratory measurements \citep{Zrlines}. We used three \ion{Nb}{I} lines located at 5189.186~\AA, 5271.524~\AA\ and 5350.722~\AA\ to derive the Nb abundances in the sample stars. Details about the Fe, Zr and Nb lines can be found in Table~\ref{linelist}.
The Tc abundance was derived using only the Tc line at 4262.27~\AA. The other s-process abundances were also determined with the goal to obtain a full s-process abundance profile (presented in Table~\ref{abundancestrinsic}). A  systematic 
\SVE{analysis} of these abundances along with those of the other Tc-rich S stars is deferred to a forthcoming paper.

\AJ{As abundance uncertainties induced by the uncertainties on the atmospheric parameters}, we used those of V915~Aql from \cite{shreeya1} since the atmospheric parameters of V915~Aql (\teff = 3400 K, \logg = 0.0, [Fe/H] = -0.5, C/O = 0.75, [s/Fe] = 0.0) are appropriate matches 
\SVE{for} 
those of our sample stars.  
We computed the total abundance uncertainty ($\sigma_{\rm[X/Fe]}$ in Table~\ref{abundancestrinsic})
by 
quadratically  adding the elemental
standard deviation 
due to line-to-line scatter, \AJ{the atmospheric parameter-induced uncertainties as estimated  for  V915~Aql},  and  a   0.1~dex uncertainty related  
to the continuum placement. When only one spectral line was used to derive the abundance of a given element, a standard line-to-line scatter of 0.1~dex was adopted. The 
abundances and 
their uncertainties are listed in Table~\ref{abundancestrinsic}.


\section{Location of the bitrinsic S stars in the HR diagram}\label{Sect:HRD}
\begin{figure}[hbt!]
\begin{centering}
    \includegraphics[scale=0.45,trim={1cm 0cm 1cm 1cm}]{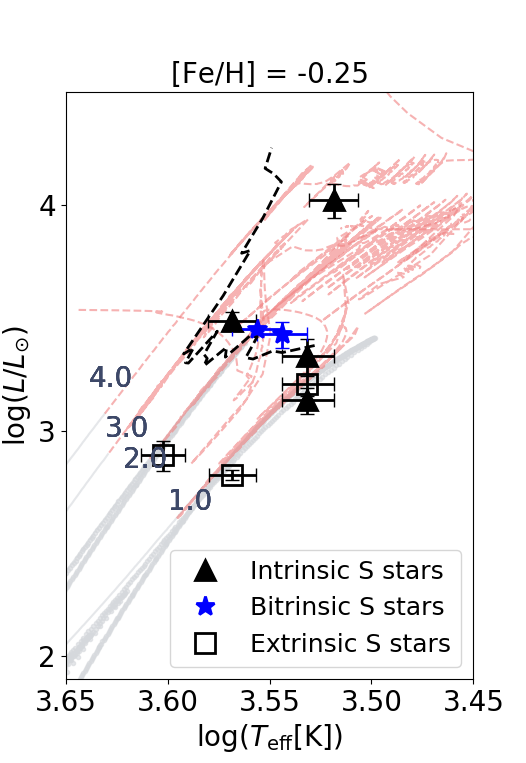}
    \caption{\label{HRD}HR diagram of intrinsic (filled triangles) S stars from \cite{shreeya1,shreeya2}, bitrinsic S stars (blue stars) and extrinsic (open squares) S stars from \cite{shreeya1} along with the STAREVOL evolutionary tracks corresponding to the closest grid metallicity. 
    The red giant branch is represented by the grey dots, whereas the red dashed line corresponds to the AGB tracks. The black dashed line marks the onset of TDU.
    }
\end{centering}
\end{figure}
In order to infer the evolutionary status of S stars, we locate them in the HR diagram 
using the effective temperature derived as described in Sect.~\ref{Sect:atmosphericparameters} and luminosities derived using the Gaia DR2 parallaxes. We compare these positions in the HR diagram with the STAREVOL evolutionary tracks of the corresponding metallicities (Fig.~\ref{HRD}). \omiori{} and \BD{} are both located 
above the black dashed line that marks the predicted onset of TDU in the HR diagram. This confirms that they are 
TP-AGB stars. 
The fact that they are not much evolved on the TP-AGB is consistent
with their moderate s-process abundances, relatively low C/O ratios, and absence of infrared excesses.

\section{Comparison with STAREVOL nucleosynthesis predictions}\label{SECT:STAREVOLpredictions}
\begin{figure}
    \centering
    \includegraphics[scale=0.47]{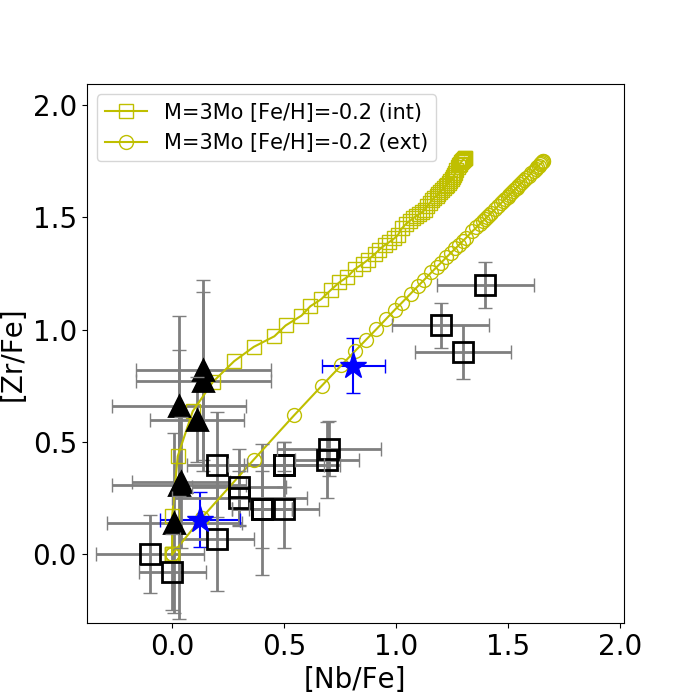}
    \caption{[Zr/Fe] vs [Nb/Fe] ratios for the sample stars (blue star symbols), as compared with intrinsic (triangles) and extrinsic (open squares) S stars from \cite{shreeya1, shreeya2}. The yellow-green curves are the nucleosynthesis predictions from STAREVOL for 3~\msun, [Fe/H] = -0.2 models. Each symbol along these curves represents a thermal pulse. The 2 \msun~models completely overlap the 3 \msun~ones but reach only up to [Zr/Fe]~$\sim$1.3.}
    \label{ZrvsNbfigure}
\end{figure}

\begin{figure}
    \centering
    \includegraphics[scale=0.44, trim={1cm 1cm 2cm 1cm}]{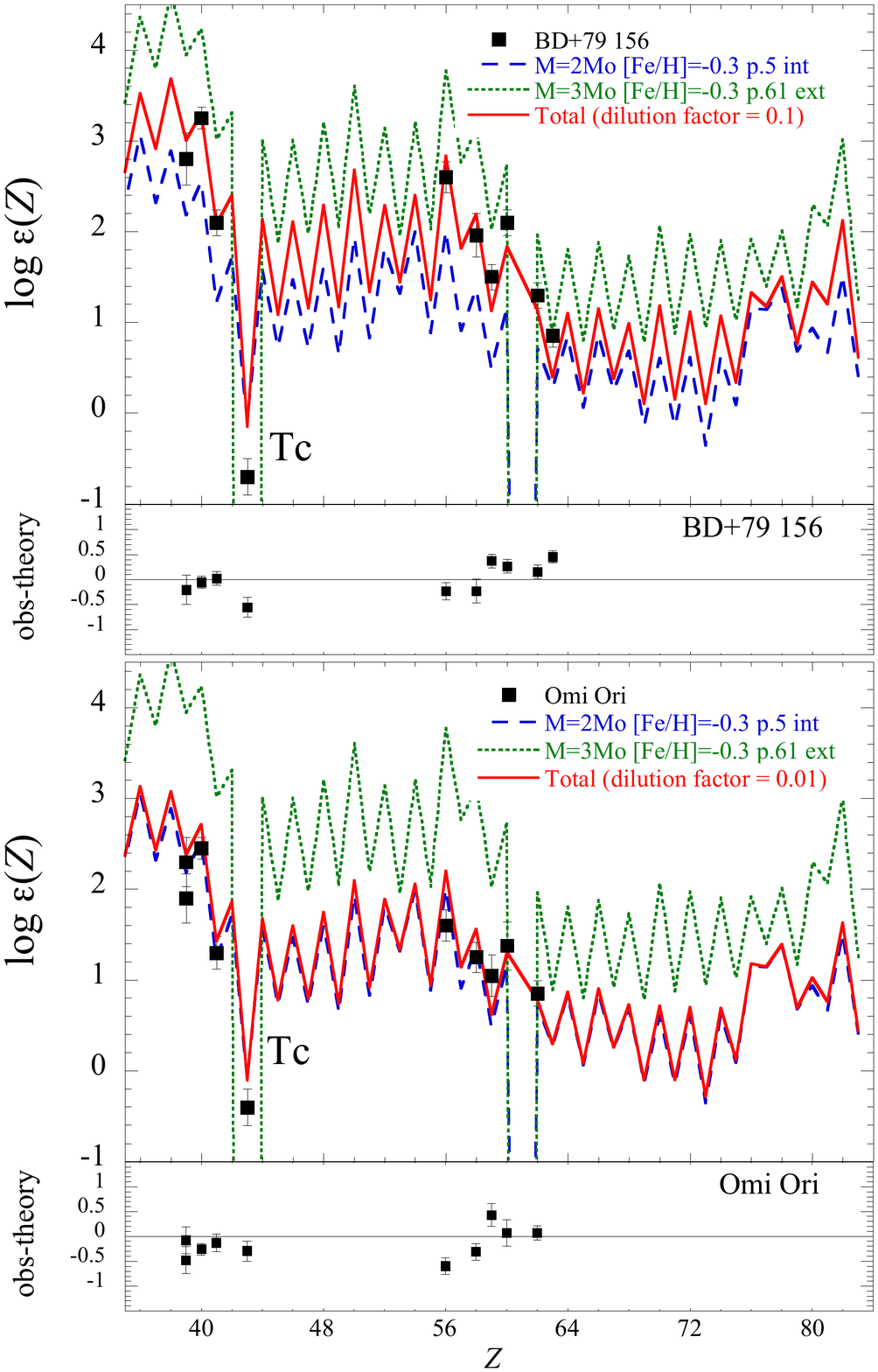}
    \caption{\label{predictionsabundprofile} Comparison of the observed and predicted surface abundances for \BD~(top panel) and \omiori{} (bottom panel) in unit of log~$\epsilon$. The best fit (solid red curves) results from the dilution of the yields from a 3~\msun{}, [Fe/H]~=~-0.3 model within the envelope of a 2~\msun\ [Fe/H]~=~-0.3 intrinsic AGB star after it has experienced 5 thermal pulses. In both panels, "p.5 int"  means that 5 pulses were involved in the intrinsic contribution whereas "p61 ext" means that 61 pulses occurred in the companion star (considering mass transfer at the end of the TP-AGB phase of the companion).}
\end{figure}

The Tc and Nb abundances are inversely correlated and a clear segregation between intrinsic and extrinsic S stars  is expected in the Zr-Nb plane, as seen in
Fig.~\ref{ZrvsNbfigure}.

We now compare the measured abundances with the nucleosynthesis predictions from the STAREVOL code \citep{siess2008, goriely&siess}. The partial mixing of protons in the C-rich layers responsible for the s-process nucleosynthesis is described by a diffusive equation with the mixing parameters $f_{\rm over} = 0.1$, $D_{\rm min} = 10^9$~cm$^2$~s$^{-1}$ and $p = 5$, where $f_{\rm over}$ controls the extent of the mixing, $D_{\rm min}$ the value of the diffusion coefficient at the innermost boundary of the diffusive region, and $p$ is an additional free parameter describing the shape of the diffusion profile. Other sets of mixing parameters could be considered \citep{shreeya2}. All details can be found in \citet{goriely&siess} and references therein.

In Fig.~\ref{ZrvsNbfigure}, we present the predictions for intrinsic and extrinsic stars with masses 2 and 3~\msun, and [Fe/H] = 0.0 and -0.2. Our sample stars \BD{} and \omiori{} are both clearly Tc-rich. According to the Nb/Tc anti-correlation, they should thus not be enriched in Nb, contrarily to what is observed in the ([Zr/Fe], [Nb/Fe]) plane (Fig.~\ref{ZrvsNbfigure}). \BD{} is found within the extrinsic stars  ([Zr/Fe]=0.84, [Nb/Fe]=0.81), along the predicted theoretical trend for extrinsic S stars. Its [Zr/Nb] ratio is 
close to unity, which indicates that $^{93}$Zr has decayed and that a mass-transfer episode from a former AGB star has taken place. However, because \BD{} is as well enriched in Tc, it must currently also be an AGB star. The bitrinsic star \BD{} thus constitutes an example of dual-enrichment events (one extrinsic, one intrinsic).

In \omiori,  Nb is overabundant but to a much lesser extent ([Nb/Fe]=0.12, [Zr/Fe]=0.15). Its position in Fig.~\ref{ZrvsNbfigure} is less conclusive. However, its [Zr/Nb] ratio is close to unity as in the case of \BD. Besides, a WD companion has been detected for \omiori{} \citep[Sect.~\ref{Sect:binarity} and][]{OmiOriAke}, strengthening the case for it being a bitrinsic object. 

\LS{
To account for the surface composition of \BD, we computed the evolution of a  2~\msun\ star with  metallicity [Fe/H]~=~-~0.3 representative of the observed star. This model will account for the production of Tc. To account for the Nb overabundance, we then assume it to be due to the $^{93}$Zr pollution from a former 3~\msun{} AGB companion, that is now a dim WD.  The composition of the current star is then the result from the enrichment coming from its companion through mass transfer as well as from its own TDU episodes. By adjusting the amount of material diluted in the envelope of the 2~\msun{} star from its companion and the intrinsic surface enrichment through TDU episodes ({\it i.e.} the number of thermal pulses encountered so far), we can reproduce fairly well the observed abundances, as shown in  Fig.~\ref{predictionsabundprofile}. 
\SVE{Within this non-unique parameter set (describing mixing and stellar masses), a non-negligible 10\% dilution factor from \BD's companion wind is needed to explain the high Nb abundance found at its surface.}
\SVE{We conclude that for} 
\BD, a double pollution event is necessary to explain the abundance pattern.  It is found not only to reproduce the measured Tc and Nb abundances but also those of light and heavy s-process elements. This conclusion holds independently of the uncertainties affecting the mixing parameters in the description of the diffusion processes. No other elements than Nb and Tc can unambiguously point out towards such a double pollution mechanism.}

Concerning \omiori{}, this system has been considered as a prototype bitrinsic star for a very long time \citep{OmiOriAke,jorissen1993, jorissen2019}. As shown in the bottom panel of Fig.~\ref{predictionsabundprofile} where we compare the abundance profile of \omiori{} with the predictions of intrinsic and extrinsic stars, a pollution of s-process elements by a binary companion may not be needed because of the slight overabundance in Nb. Hence \omiori{} is actually a star with only a marginal bitrinsic character. We support the claims of \cite{OmiOriAke} that, though the existence of a WD companion is very clear for \omiori, there are no signatures in the chemical abundances that would need an explanation other than the occurrence of intrinsic s-process nucleosynthesis followed by TDUs on the AGB. We thus regard \omiori{} as a Tc-rich S star with a WD companion but with no clear evidence for a large
pollution coming from the AGB progenitor of this companion.

\section{Conclusions}\label{Sect:Conclusions}

S stars of a new type have been uncovered, bearing characteristics of both intrinsic and extrinsic S stars, and referred to as "bitrinsic" stars. 
The intriguing hybrid classification as both intrinsic (because of Tc detection) and extrinsic (because of Nb overabundance and WD companion detection) can be understood if we identify these stars with objects formerly polluted by an AGB companion, and now evolving on the TP-AGB.
Though such objects were expected to exist, none had been identified until now because of the lack of appropriate diagnostics. 
Thanks to a careful analysis of S-type stars involving atmospheric parameter and abundance determination using HERMES high-resolution spectra, we could \LS{identify one clear  case} (\BD{}) of such doubly-polluted objects, and one tentative case (\omiori{}).

These objects are interesting in several respects. 
 First, if an appropriate statistics of bitrinsic stars were available, this would put interesting constraints on binary-synthesis evolution models, because the number of bitrinsic star depends, among other  things, on the radius evolution during the RGB and AGB phases, which will determine whether/when a Roche-lobe overflow will occur, causing the disappearance of systems that would have become bitrinsic objects \citep{Escorza-2020}. A second constraint concerns the component masses: in order to exhibit s-process overabundances at their surface, AGB stars must have an initial mass larger than 1~M$_\odot$ \citep[as recently shown by][]{shreeya2}  and smaller than $4-5$~M$_\odot$, above which the radiative s-process production is significantly reduced \citep{Goriely-2004}. Therefore, in the case of bitrinsic systems, both the primary {\it and} secondary components should satisfy this condition (along with $M_1 > M_2$).
Third, spectroscopic binaries involving AGB stars are difficult to find because their radial-velocity variations are blurred by strong imprint from envelope pulsations. 
Tc and Nb diagnostics thus offer a direct spectroscopic way to prove the binarity of an AGB star, without resorting to a time-consuming radial-velocity monitoring.
\SVE{Fourth, bitrinsic stars should exist among other stellar classes as well, and in the future they should be sought for among carbon stars and possibly among lower-metallicity carbon-enriched objects. 
}

\begin{acknowledgements}
      This research has been funded by the Belgian Science Policy Office under contract BR/143/A2/STARLAB. 
      S.V.E. thanks {\it Fondation ULB} for its support.
      Based on observations obtained with the HERMES spectrograph, which is supported by the Research Foundation - Flanders (FWO), Belgium, the Research Council of KU Leuven, Belgium, the \textit{Fonds National de la Recherche Scientifique} (F.R.S.-FNRS), Belgium, the Royal Observatory of Belgium, the \textit{Observatoire de Gen\`eve}, Switzerland and the \textit{Th\"{u}ringer Landessternwarte Tautenburg}, Germany. LS \& SG are senior FNRS research associates. AE acknowledges the support of FWO under contract ZKD1501-00-W01.
\end{acknowledgements}

\bibliographystyle{aa}
\bibliography{aa}

\begin{appendix}
\section{Spectral energy distributions of the sample stars}

\begin{figure}[ht]
    \includegraphics[scale=0.409]{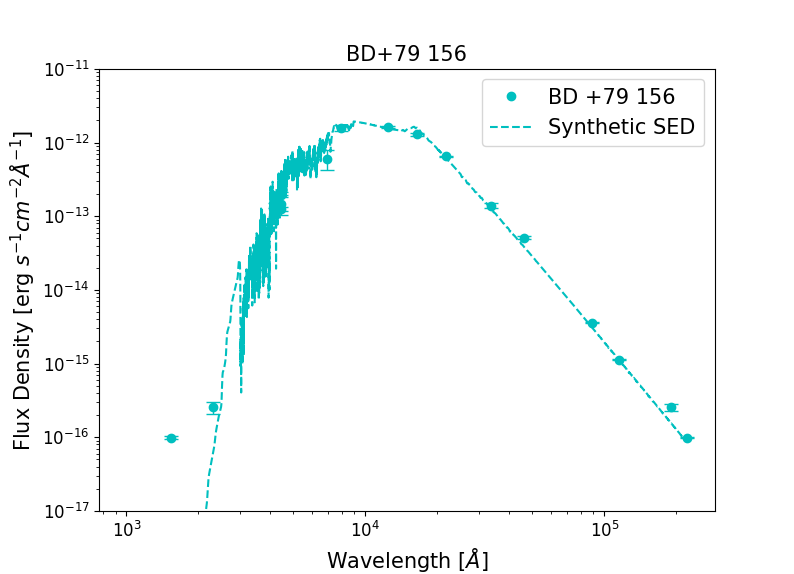}
    \includegraphics[scale=0.41]{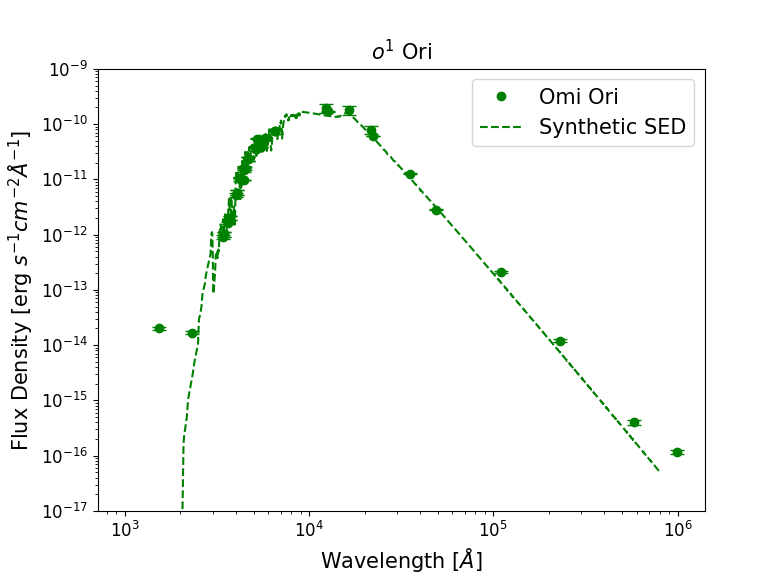}
    \caption{Photometric observations of \BD{} (top panel) and \omiori{} (bottom panel) compared with the synthetic spectral energy distribution generated using the MARCS model atmospheres with the parameters listed in Table~\ref{basicdataandparameters}.
    All photometric observations were retrieved from SIMBAD \citep{simbad}. Note in the two stars the evident UV excess in the bluest flux point. }
    \label{SEDs}
\end{figure}

\begin{table*}
\caption{Elemental abundances of the two bitrinsic candidates, along with the standard deviation due to line-to-line scatter. Solar abundances (third column) are from \cite{asplund2009}. The  column labelled $N$ lists the number of lines used to derive the corresponding elemental abundance. The $\sigma_{\rm [X/Fe]}$ column lists the total uncertainty on the abundance calculated using the method described in Sect.~\ref{Sect:abundances}.}
\label{abundancestrinsic}
\centering
\begin{tabular}{lrc | lrccl | lrccl }
\hline
 & & &\multicolumn{5}{c}{\omiori}&\multicolumn{5}{c}{\BD} \\
 \hline
      & Z  & $\log {\epsilon}_\odot$ & $\log \epsilon$ & $N$   & [X/H] & [X/Fe] & $\sigma_{\rm[X/Fe]}$  & $\log \epsilon$ & $N$   & [X/H] & [X/Fe] & $\sigma_{\rm[X/Fe]}$ \\
      \hline
C   & 6  & 8.43  & 8.23 &    & -0.19    & 0.09       &   0.10          & 8.36 &    & -0.07    & 0.09 & 0.17 \\
N   & 7  & 7.83  & 8.20 $\pm$ 0.13  &    & 0.37      & 0.65      & 0.65        & 8.70 $\pm$ 0.13 &    & 0.87      & 1.04     & 0.65  \\
O     & 8  & 8.69  & 8.36  &    & -0.33     & -0.05     &             & 8.66  &    & -0.03     & 0.14      &         \\
Fe    & 26 & 7.50   & 7.21 $\pm$ 0.19  & 13 & -0.28    & -          & 0.23        & 7.33 $\pm$ 0.12  & 13 & -0.17   & -       & 0.18   \\
Y I   & 39 & 2.21  & 1.90 $\pm$ 0.00  & 1  & -0.31     &  -0.02    & 0.27        & 2.80 $\pm$ 0.00 & 1  & 0.59     & 0.76     & 0.29  \\
Y II  & 39 & 2.21  & 2.30 $\pm$ 0.00   & 1  & 0.09      & 0.37      & 0.27  & -     & -   & -& -   & - \\
Zr I  & 40 & 2.58  & 2.45 $\pm$ 0.07  & 2  & -0.13     & 0.15      & 0.12        & 3.25 $\pm$ 0.07 & 2  & 0.67      & 0.84      & 0.12  \\
Nb I  & 41 & 1.46  & 1.30 $\pm$ 0.15  & 3  & -0.16     & 0.12      & 0.18        & 2.10 $\pm$ 0.00   & 3  & 0.64     & 0.81  & 0.14   \\
Tc  & 43  & & -0.4 $\pm$ 0.00 & 1 & & & 0.17 & -0.7 $\pm$ 0.00 & 1 & & & 0.17\\
Ba I  & 56 & 2.18  & 1.60 $\pm$ 0.00   & 1  & -0.58     & -0.29     & 0.17        & 2.60 $\pm$ 0.00   & 1  & 0.42      & 0.59    & 0.17 \\
Ce II & 58 & 1.58  & 1.25 $\pm$ 0.28  & 6  & -0.33     & -0.04     & 0.17        & 1.96 $\pm$ 0.28  & 7  & 0.38      & 0.55     & 0.24 \\
Pr II & 59 & 0.72  & 1.05 $\pm$ 0.21  & 2  & 0.33      & 0.61      & 0.23        & 1.50 $\pm$ 0.00 & 1   & 0.78 & 0.95   &0.14 \\
Nd II & 60 & 1.42  & 1.37 $\pm$ 0.15 & 5  & -0.04    & 0.24     & 0.26        & 2.10 $\pm$ 0.00   & 2  & 0.68      & 0.85     & 0.14    \\
Sm II & 62 & 0.96  & 0.85 $\pm$ 0.10  & 2  & -0.11     & 0.17     & 0.14        & 1.30 $\pm$ 0.00 & 1 & 0.34  & 0.51 & 0.14 \\
Eu II & 63 & 0.52  & -     & -  & -& -   & -            & 0.85 $\pm$ 0.00  & 1  & 0.33      & 0.50        & 0.12  \\
\hline \\       
\end{tabular}
\end{table*}

\begin{table*}
\centering
\caption{Atomic lines used in this study.}
\label{linelist}
\begin{tabular}{c c c c c}

\hline
Species& $\lambda$ [\AA] & $\chi$ [eV] & $\log gf$ & Reference \\
\hline
Fe I & 7389.398 & 4.301 & -0.460 & \cite{K07} \\
 & 7418.667 & 4.143 & -1.376 & \cite{BWL} \\
 & 7443.022 & 4.186 & -1.820 & \cite{MFW} \\
 & 7461.263 & 5.507 & -3.059 & \cite{K07} \\
 & 7498.530 & 4.143 & -2.250 & \cite{MFW} \\
 & 7540.430 & 2.727 & -3.850 & \cite{MFW} \\
 & 7568.899 & 4.283 & -0.773 & \cite{K07} \\
 & 7583.787 & 3.018 &  -1.885 &    \cite{BWL}  \\
 & 7586.018 & 4.313 & -0.458 & \cite{K07} \\
 & 7832.196 & 4.435 & 0.111 & \cite{K07} \\
 & 8108.320 & 2.728 & -3.898 & \cite{K07} \\
 & 8698.706 & 2.990 & -3.452 & \cite{K07} \\
 & 8699.454 & 4.955 & -0.380 & \cite{NS} \\
 & 8710.404 & 5.742 & -5.156 & \cite{K07} \\
 & 8729.144 & 3.415 & -2.871 & \cite{K07} \\
 Zr I & 7819.374 & 1.822 & -0.380 & \cite{Zrlines} \\
 & 7849.365 & 0.687 & -1.300 & \cite{Zrlines}\\
Nb I & 5189.186 & 0.130 & -1.394 & \cite{DLa} \\
 & 5271.524 & 0.142 & -1.240 & \cite{DLa} \\
 & 5350.722 & 0.267 & -0.862 & \cite{DLa} \\
\hline

\end{tabular}

\end{table*}

\end{appendix}

\end{document}